\documentclass{ifacconf}

\usepackage{graphicx}      
\usepackage{natbib}        

\usepackage{multicol}
\usepackage{amsmath,amssymb}
\usepackage{amsfonts}
\usepackage{textcomp}
\usepackage{psfrag}
\usepackage{multimedia}
\usepackage{fancybox}

\newcommand{\matr}[2]{\left[\begin{array}{#1}#2\end{array}\right]}
\newcommand{\E}[2]{\mathbb{E}_{#1}\left[#2\right]}

\newcommand{\vect}[1]{{\ensuremath{\boldsymbol{\mathrm{#1}}}}}

\newtheorem{Remark}[thm]{Remark}

\usepackage[usenames,dvipsnames]{color}
\definecolor{wheat}{rgb}{0.96,0.87,0.70}
\definecolor{mario}{rgb}{0.8,0.8,1}
\definecolor{seb}{rgb}{0.8,1,0.8}

\begin{document}
\begin{frontmatter}

\title{Reinforcement Learning Based on Real-Time Iteration NMPC}


\author[First]{Mario Zanon} 
\author[Second]{Vyacheslav Kungurtsev} 
\author[Third]{S\'ebastien Gros} 

\address[First]{IMT School for Advanced Studies Lucca, Italy}
\address[Second]{Czech Technical University in Prague, Czech Republic}
\address[Third]{Norwegian University of Technology, NTNU}

\begin{abstract}                
	Reinforcement Learning (RL) has proven a stunning ability to learn optimal policies from data without any prior knowledge on the process. The main drawback of RL is that it is typically very difficult to guarantee stability and safety. On the other hand, Nonlinear Model Predictive Control (NMPC) is an advanced model-based control technique which does guarantee safety and stability, but only yields optimality for the nominal model. Therefore, it has been recently proposed to use NMPC as a function approximator within RL. 
	While the ability of this approach to yield good performance has been demonstrated, the main drawback hindering its applicability is related to the computational burden of NMPC, which has to be solved to full convergence. In practice, however, computationally efficient algorithms such as the Real-Time Iteration (RTI) scheme are deployed in order to return an approximate NMPC solution in very short time. In this paper we bridge this gap by extending the existing theoretical framework to also cover RL based on RTI NMPC.
	We demonstrate the effectiveness of this new RL approach with a nontrivial example modeling a challenging nonlinear system subject to stochastic perturbations with the objective of optimizing an economic cost.
\end{abstract}
 
\end{frontmatter}

\section{Introduction}\label{sec:intro}
Reinforcement Learning (RL) is a powerful data-driven technique which aims at optimizing the performance of a Markov Decision Process (MDP), typically without relying on a model of the state transition probability. Most RL methods rely purely on the observed stage cost and state transition in order to estimate and optimize the closed-loop performance. The potential of RL has been recently demonstrated by several successful implementations including, e.g., robots learning to walk or fly~\citep{Wang2012,Abbeel2007}, or computers beating chess and go champions~\citep{Silver2016}.

Some RL methods optimize the performance indirectly, by parametrizing the action-value function and learning a good approximation of the optimal action-value function underlying the MDP. The optimal policy is then obtained by optimizing the action-value function with respect to the action. Other RL methods, instead, directly parametrize the policy and estimate its gradient in order to optimize the policy parameter. Both direct and indirect approaches typically rely on derivative-based stochastic optimization, such that the function approximator must be differentiable in the parameter.

Deep Neural Networks (DNN) are a common choice to support the parametrization of either the action-value function or the policy. While DNN can be very effective in practice, they pose difficulties in the analysis of closed-loop stability or in imposing hard constraints on the evolution of the state of the real system. An alternative to DNNs has been proposed in~\citep{Gros2020}, where it has been suggested to use nonlinear Model Predictive Control (MPC) as a function approximator. Strategies to provide strict constraint satisfaction guarantees have been investigated in~\citep{Gros2020c} by using projection approaches and in~\citep{Zanon2019b,Gros2020d} by using robust MPC schemes. The use of mixed-integer MPC formulations has been investigated in~\citep{Gros2020b}. Furthermore, the use of MPC as a function approximator gives one the possibility to easily introduce any available information on the system dynamics. 
Which MPC parameter to adapt using RL is a design decision and can include, e.g., the cost Hessian and gradient, some model or constraint parameter, etc.

One of the main drawbacks of MPC, however, is the need to solve an optimal control problem in real time. While this issue is less significant for linear systems, for nonlinear MPC (NMPC) tailored algorithms are required in order to limit the computational burden and guarantee real-time feasibility for many applications of interest. In particular, the real-time iteration (RTI) scheme~\citep{Diehl2005} obtains the complexity reduction by solving an individual quadratic program (QP), i.e., one step of sequential quadratic programming (SQP). By exploiting the similarity between two consecutive MPC problems, the solution of the first can be used to construct a good initial guess for the second. As a result, each OCP solve is reduced to one (or a small number) of QP solves, which can be performed quickly using contemporary solvers.

The use of RTI-based NMPC as a function approximator for RL is a natural means of making MPC-based RL applicable in real time to a wide class of systems. In order to be able to deploy standard RL techniques, one must be able to compute the sensitivities of RTI-based NMPC with respect to the RL parameter. The necessary sensitivity analysis for fully converged NMPC has been proposed in~\citep{Gros2020}, by relying on results on parametric optimization. However, the analysis of~\citep{Gros2020} does not apply to the RTI scheme, since it assumes that the NMPC problem is solved to full convergence. In this paper, we set the theoretical foundations which allow one to deploy the RTI scheme within RL and demonstrate the effectiveness of our approach in simulations by minimizing the economic cost of a nonlinear system subject to stochastic perturbations.

The paper is structured as follows. In Section~\ref{sec:background}, we provide some background on Q-learning and
OCPs as function approximation. In Section~\ref{sec:mpc} we present nonlinear MPC and the RTI scheme; we provide the sensitivity analysis for the RTI scheme in Section~\ref{sec:sensitivity}; and in Section~\ref{sec:rl_mpc} we discuss the adaptation of classic RL methods when deploying them in combination with NMPC. 
We report on some numerical simulations in Section~\ref{sec:simulations}. In Section~\ref{sec:conclusions} we conclude the paper and outline future research directions.

\section{Reinforcement Learning Background}\label{sec:background}

Consider a system whose dynamics are described by a Markov Process (MP) with with continuous state $\vect{s}$ and action (or control) $\vect{a}$, and state transition $\vect{s},\vect{a}\rightarrow\vect{s}_+$ described by probability density
\begin{align}
	\label{eq:state_transition}
	\mathbb{P}\left[\vect{s}_{+}\,|\,\vect{s},\vect{a}\right].
\end{align}
Note that often system~\eqref{eq:state_transition} is written as
	$\vect{s}_+ = \vect{f}(\vect{s},\vect{a},\vect{w}), $
with $\vect{w}$ stochastic process noise described by probability density $\mathbb{P}_\vect{w}$.
We Associate with system~\eqref{eq:state_transition} the stage cost $\ell(\vect{s},\vect{a})$ and the discount factor $\gamma\in[0,1]$ to 
define a Markov Decision Process (MDP).

Consider a deterministic policy delivering the control input $\vect{a}=\vect{\pi}(\vect{s})$, resulting in state distribution $\tau^\vect{\pi}$. Reinforcement Learning aims at finding the best policy $\vect{\pi}_\star$, i.e., at solving
\begin{align}
	\label{eq:rl_problem}
	\vect{\pi}_\star:=\arg\min_{\vect{\pi}}\ J(\vect{\pi}) & := \E{\tau^{\vect{\pi}}}{\sum_{k=0}^\infty \gamma^k \ell\left (\vect{s}_k,\vect{\pi}\left (\vect{s}_k\right )\right )}.
\end{align}
Important quantities in RL are the  action-value function $Q^\star(s,a)$ and value function $V^\star(s)$ associated with the optimal policy $\pi^\star\left(s\right)$, defined by the Bellman equations:
\begin{subequations}
	\label{eq:Bellman}
	\begin{align}
	Q^\star\left(s,a\right) &= \ell\left(s,a\right) + \gamma \mathbb{E}\left[V^\star(s_+)\,|\, s,a\right], \label{eq:Bellman1}\\
	V^\star\left(s\right) &= Q^\star\left(s,\pi_\star\left(s\right)\right) = \min_{a}\, Q^\star\left(s,a\right). \label{eq:Bellman2}
	\end{align}
\end{subequations}

Several RL algorithms have been proposed in the literature and two popular approaches are $Q$-learning and policy gradient methods~\citep{Sutton2018}. We provide next a brief description of these two approaches.

\subsection{$Q$-learning}
$Q$-learning parametrizes the action value function as $Q_\vect{\theta}(\vect{s},\vect{a})$, where $\vect{\theta}$ is a vector of parameters whose values have to be learned, and aims at minimizing $\| Q^\star(\vect{s},\vect{a}) - Q_\vect{\theta} (\vect{s},\vect{a}) \|_2^2$.
Standard algorithms rely on the recursive parameter update~\citep{Sutton2018}
\begin{subequations}
	\label{eq:Qlearning}
	\begin{align}
		\hspace{-0.8em}\delta_k &= \ell(\vect{s}_k,\vect{a}_k) + \gamma\,\min_{\vect{a}_{k+1}} Q_\vect{\theta}(\vect{s}_{k+1},\vect{a}_{k+1}) - Q_\vect{\theta}(\vect{s}_k,\vect{a}_k),\hspace{-0.3em} \label{eq:TDError}\\
		\hspace{-0.8em}\vect{\theta} &\leftarrow \vect{\theta} + \alpha\delta_k\nabla_\vect{\theta} Q_\vect{\theta}(\vect{s}_k,\vect{a}_k). \label{eq:Qlearning:Update}
	\end{align}
\end{subequations}
$Q$-learning has been successfully applied in, e.g.,~\citep{Watkins1989,Mnih2015,Theocharous2015}.

\subsection{Actor-Critic Policy Gradient}
Policy gradient approaches parametrize the policy as $\vect{\pi}_\vect{\theta}$ and directly aim at approximately solving~\eqref{eq:rl_problem} as
\begin{align}
	\label{eq:policy_gradient}
	\vect{\theta}_\star = \arg\min_{\vect{\theta}}\ J(\vect{\pi}_\vect{\theta}).
\end{align}
This typically involves some form of stochastic descent algorithm where, in a deterministic policy gradient framework based on actor-critic methods, we have
\begin{subequations}
	\label{eq:pg_update}
	\begin{align}
		\nabla_\vect{\theta} J(\vect{\pi}_\vect{\theta}) &= \E{\vect{\pi}_\vect{\theta}}{\nabla_\vect{\theta} \vect{\pi}_\vect{\theta} \nabla_\vect{a}A_{\vect{\pi}_\vect{\theta}}}, \\
		\vect{\theta} & \leftarrow \vect{\theta} + \alpha \nabla_\vect{\theta} J(\vect{\pi}_\vect{\theta}),
	\end{align}
\end{subequations}
with advantage function $A_{\vect{\pi}_\vect{\theta}}(\vect{s},\vect{a})=Q_{\vect{\pi}_\vect{\theta}}(\vect{s},\vect{a})-V_{\vect{\pi}_\vect{\theta}}(\vect{s})$.

\subsection{Function Approximation based on MPC}

The use of NMPC to parametrize the action-value function has been first investigated in~\citep{Gros2020}, where it has been proven that NMPC is a universal function approximator in the sense that it can be used to support the action value function $Q_\vect{\theta}$, value function $V_\vect{\theta}$ and corresponding policy $\vect{\pi}_\vect{\theta}$ at once. We remark that NMPC is a nonlinear function approximator which typically requires to be formulated using a positive-definite cost. In general, however, the cost to be optimized by RL can be indefinite.


As proven in~\citep{Gros2020} this does not limit the applicability of NMPC in the RL context, provided that an initial cost term is added to the problem formulation to perform a so-called \emph{cost rotation}. This rotation makes it possible to approximate an indefinite $Q$ while using a positive-definite cost in NMPC.


We ought to stress that the RL-NMPC framework proposed in~\citep{Gros2020} and further investigated in~\citep{Zanon2019,Zanon2019b,Gros2020d,Gros2020b} does not provide strict stability guarantees. Such guarantees would be given in case the discount factor were $\gamma=1$ and suitable terminal conditions were formulated~\citep{Rawlings2009b,Grune2011}. Clearly, one can formulate the NMPC problem by using $\gamma=1$ even when the MDP has $\gamma <1$. However, the ability of NMPC to approximate  $Q_\star, V_\star, \vect{\pi}_\star$ has not been thoroughly investigated. This is the subject of ongoing research, but beyond the scope of this paper.

\section{NMPC}
\label{sec:mpc}



In this paper, we consider function approximators based on OCP parametrized by $\vect{\theta}$ of the form
\begin{subequations}
	\label{eq:param_nmpc}
	\begin{align}
	Q_\vect{\theta}(\vect{s},\vect{a}) =  \min_{\vect{z}}\ \ & \lambda_\vect{\theta}(\vect{s})+ \gamma^N V^\mathrm{f}_\vect{\theta}(\vect{x}_N) + \sum_{k=0}^{N-1} \gamma^k \ell_\vect{\theta}(\vect{x}_k,\vect{u}_k) \label{eq:param_nmpc:cost}\\
	\mathrm{s.t.} \ \ & \vect{x}_0 = \vect{s}, \quad \vect{u}_0=\vect{a}, \label{eq:param_nmpc:initial}\\
	&\vect{x}_{k+1} = \vect{f}_\vect{\theta}\left(\vect{x}_k,\vect{u}_k\right), \label{eq:param_nmpc:dynamics}\\
	& \vect{g}\left(\vect{u}_k\right) \leq 0, \label{eq:param_nmpc:input_const} \\
	& \vect{h}_\vect{\theta}\left(\vect{x}_k,\vect{u}_k\right) \leq 0,\quad \vect{h}^\mathrm{f}_\vect{\theta}(\vect{x}_N) \leq 0, \label{eq:Const:Relaxation} 
	\end{align}
\end{subequations}
where $\vect{z}=(\vect{x}_0,\vect{u}_0,\ldots,\vect{x}_N)$. The stage and terminal cost $\ell_\vect{\theta}, V^\mathrm{f}_\vect{\theta}$, the system dynamics and constraints $\vect{f}_\vect{\theta},\vect{h}_\vect{\theta},\vect{h}^\mathrm{f}_\vect{\theta}$ and the initial cost $\lambda_\vect{\theta}$ are parametric functions of $\vect{\theta}$. Note that in MPC the initial constraint~\eqref{eq:param_nmpc:initial} typically only involves the state, i.e., $\vect{u}_0=\vect{a}$ is not present, since the goal is to compute an optimal policy. The policy $\vect{\pi}_\vect{\theta}(\vect{s})$ and value function $V_\vect{\theta}(\vect{s})$ are obtained by solving Problem~\eqref{eq:param_nmpc} with constraint $\vect{u}_0=\vect{a}$ removed. This is fully equivalent to
\begin{align}
	\label{eq:param_nmpc_s}
	\hspace{-0.5em} \vect{\pi}_\vect{\theta}(\vect{s}) = \mathrm{arg}\min_\vect{a}\, Q_\vect{\theta}(\vect{s},\vect{a}), \quad V_\vect{\theta}(\vect{s}) = \min_\vect{a}\, Q_\vect{\theta}(\vect{s},\vect{a}).
\end{align}

In standard NMPC, parameter $\vect{\theta}$ is typically considered as follows. The system dynamics and constraints $\vect{f}_\vect{\theta},\vect{h}_\vect{\theta},\vect{h}^\mathrm{f}_\vect{\theta}$, are derived as mathematical models of the physical process that needs to be controlled. System identification techniques are deployed to compute parameter $\vect{\theta}$ such that the model predictions fit experimental data sufficiently accurately. Concerning the cost, instead, parameter $\vect{\theta}$ is a tuning parameter which should be chosen by the control engineer to ensure that the closed-loop system performance is satisfactory, e.g., in terms of disturbance rejection and asymptotic stability. Alternatively, if---similarly to RL---a clear objective is available, so-called \emph{economic MPC} schemes directly optimize the prescribed objective~\citep{Rawlings2009b,Grune2011}.

In order to address feasibility issues in Problem~\eqref{eq:param_nmpc} we propose to adopt an exact relaxation of state-dependent constraints, as proposed in~\citep{Scokaert1999a} and used in the context of RL in~\citep{Gros2020,Zanon2019,Zanon2019b}. 

\subsection{Real-Time NMPC}
Problem~\eqref{eq:param_nmpc} is a parametric nonlinear programming problem, the solution of which can be computationally demanding. 
Several approaches have been proposed in order to solve Problem~\eqref{eq:param_nmpc} approximately but quickly, e.g., the Real-Time Iteration (RTI) scheme~\citep{Diehl2002b}, the Advanced Step NMPC Controller~\citep{Zavala2009} and the continuation/GMRES approach~\citep{Ohtsuka2004}. All these approaches are based on pathfollowing techniques for parametric NLPs, where the parameter of interest is the initial state $\vect{s}$, and essentially rely on the (approximate) solution at the previous time instant to build a good initial guess for the problem at the current time in a predictor-corrector framework. For the sake of simplicity, we focus on the RTI scheme, which is based on sequential quadratic programming and operates as follows.

We write Problem~\eqref{eq:param_nmpc} in the compact form
\begin{align}
	\label{eq:nlp}
	\min_{\vect z} \ \ & F_\vect{\theta}(\vect{z}) &
	\mathrm{s.t.} \ \ &\vect{G}_\vect{\theta}(\vect{z}) =0, & \vect{H}_\vect{\theta}(\vect{z}) \leq 0,
\end{align}
with Lagrange multipliers $\vect{\xi},\vect{\upsilon}$ associated with $\vect{G},\vect{H}$ respectively.
Given an initial guess $\vect{y}^{(0)}=(\vect{z}^{(0)},\vect{\xi}^{(0)},\vect{\upsilon}^{(0)})$, Sequential Quadratic Programming (SQP) solves~\eqref{eq:nlp} by computing a suitable step size $t$ and updating 
\begin{align*}
	\vect{y}^{(i+1)} = \vect{y}^{(i)} + t \vect{y}^\mathrm{QP},
\end{align*}
where $\vect{y}^\mathrm{QP}$ is the primal-dual solution of the quadratic program (QP)
\begin{subequations}
	\label{eq:param_nlp_qp}
	\begin{align}
		\min_{\vect z} \ \ & \frac{1}{2} \vect{z}^\top L^{(i)} \vect{z} + \nabla_\vect{z} \mathcal{L}_\vect{\theta}\left (\vect{y}^{(i)}\right )^\top \vect{z} \\
		\mathrm{s.t.} \ \ & \nabla_\vect{z} \vect{G}_\vect{\theta}\left (\vect{z}^{(i)}\right )^\top \vect{z} + \vect{G}_\vect{\theta}\left (\vect{z}^{(i)}\right ) = 0, \\
		& \nabla_\vect{z} \vect{H}_\vect{\theta}\left (\vect{z}^{(i)}\right )^\top \vect{z} + \vect{H}_\vect{\theta}\left (\vect{z}^{(i)}\right ) \geq 0,
	\end{align}
\end{subequations}
with $\mathcal{L}_\vect{\theta}\left (\vect{y}^{(i)}\right )$ the Lagrangian of Problem~\eqref{eq:param_nmpc}, and $L^{(i)}$ a suitable approximation of $\nabla^2_{\vect{z}\vect{z}} \mathcal{L}_\vect{\theta}\left (\vect{y}^{(i)}\right )$.

The RTI scheme solves only one QP per time instant, uses $t=1$, and relies on the \emph{initial value embedding}, i.e., constraints~\eqref{eq:param_nmpc:initial} are not eliminated from the problem. Additionally, computations are split in a \emph{preparation phase} in which the sensitivities are evaluated and the QP KKT matrix is factorized. Once the initial state is available, the QP is solved and the control is applied to the system. All details on the RTI scheme can be found in~\citep{Diehl2005,Gros2020a} and references therein.

By considering $\vect{y}^\diamond$ as a function of $\vect{s}$, implicitly defined as the optimal primal-dual solution of Problem~\eqref{eq:param_nmpc_s}, the RTI scheme can be viewed as a pathfollowing predictor-corrector scheme. Since NMPC predicts the future evolution of the system, provided that the perturbations acting on the system are not excessively large, the state at the next time step $i$ will satisfy $\vect{s}_{i+1}\approx \vect{x}^\star_1(\vect{s}_i).$ RTI then exploits the good model prediction ability and the fast contraction of Newton-type methods to closely track the optimal solution $\vect{y}^\diamond(\vect{s})$. 
Similar considerations apply to the primal-dual solution $\vect{y}^\star$ of Problem~\eqref{eq:param_nmpc}, provided that $\|\vect{a}-\vect{\pi}_\vect{\theta}\|$ is small.

\section{Sensitivity Analysis}
\label{sec:sensitivity}

In order to be able to use NMPC as a function approximator for $Q$-learning or actor-critic methods, one needs to be able to compute the parametric sensitivities $\nabla_\vect{\theta} Q_\vect{\theta}$, $\nabla_\vect{\theta} V_\vect{\theta}$, $\nabla_\vect{\theta} \pi_\vect{\theta}$. 
In the following, we first recall the results for the case in which NMPC is solved to full convergence. Afterwards, we discuss the case of the RTI scheme, which is constructed using a similar reasoning.

Note that the parametric sensitivities of optimization problems exist under the assumption that linear independence constraint qualification and the strong second-order sufficient conditions hold~\citep{Nocedal2006,Buskens2001}. Therefore, extreme care must be taken to ensure that Problem~\eqref{eq:param_nmpc} satisfies both conditions.

\subsection{Sensitivities of Fully Converged NMPC}
We detail next how to compute the derivatives of the action-value function with respect to the parameters, in the same way as in~\citep{Gros2020}.
To this end, we define the Lagrange function underlying NMPC problem~\eqref{eq:param_nmpc} as
\begin{subequations}
	\begin{align*}
	\mathcal{L}_\vect{\theta}(\vect{y}) = \ &\lambda_\vect{\theta}(\vect{x}_0)+ \gamma^N V^\mathrm{f}_\vect{\theta}(\vect{x}_N)  + \vect{\chi}_0^\top\left(\vect{x}_0 - \vect{s}\right) + \vect{\mu}_N^\top \vect{h}^\mathrm{f}_\vect{\theta}(\vect{x}_N)\\ 
	& + \sum_{k=0}^{N-1}  \vect{\chi}_{k+1}^\top\left(\vect{f}_\vect{\theta}\left(\vect{x}_k,\vect{u}_k\right) - \vect{x}_{k+1} \right) + \vect{\nu}_k^\top \vect{g}_\theta\left(\vect{u}_k\right)\\
	&+\gamma^k \ell_\vect{\theta}(\vect{x}_k,\vect{u}_k) + \vect{\mu}_{k}^\top \vect{h}_\vect{\theta}\left(\vect{x}_k,\vect{u}_k\right) + \vect{\zeta}^\top(\vect{u}_0-\vect{a}),
	\end{align*}
\end{subequations}
where $\vect{\chi},\vect{\mu},\vect{\nu},\vect{\zeta}$ are the multipliers associated to constraints \eqref{eq:param_nmpc:initial}-\eqref{eq:Const:Relaxation} and $\vect{y}=(\vect{z},\vect{\chi},\vect{\mu},\vect{\nu},\vect{\zeta})$. Note that, for $\vect{\zeta}=0$, $\mathcal{L}_\vect{\theta}(\vect{y})$ is the Lagrange function associated to the NMPC problem defining the value function $\min_\vect{a} Q_\vect{\theta}(\vect{s},\vect{a})$.
We observe that \citep{Buskens2001}
\begin{align}
\label{eq:NMPCQgradient}
\nabla_\vect{\theta} Q_\vect{\theta}(\vect{s},\vect{a}) = \nabla_\vect{\theta} \mathcal{L}_\vect{\theta}(\vect{y}^\star)
\end{align}
holds for $\vect{y}^\star$ given by the primal-dual solution of \eqref{eq:param_nmpc}. Note that this equality holds because constraints \eqref{eq:param_nmpc:initial} are not an explicit function of $\vect{\theta}$. The gradient \eqref{eq:NMPCQgradient} is therefore straightforward to build as a by-product of solving the NMPC problem \eqref{eq:param_nmpc}. We additionally observe that
\begin{align}
	\label{eq:NMPCVgradient}
	\nabla_\vect{\theta} V_\vect{\theta}(\vect{s}) = \nabla_\vect{\theta} \min_\vect{a} Q_\vect{\theta}(\vect{s},\vect{a}) = \nabla_\vect{\theta} \mathcal{L}(\vect{y}^\diamond),
\end{align}
where $\vect{y}^\diamond$ is given by the primal-dual solution to \eqref{eq:param_nmpc_s}, i.e., \eqref{eq:param_nmpc} with constraint $\vect{u}_0=\vect{a}$ removed and $\vect{\zeta}^\diamond=0$.

The derivative of the optimal primal-dual solution with respect to the parameters is given by
\begin{align}
	\label{eq:dpolicy_dtheta}
	\nabla_{\vect{\theta}} \vect{y}^\diamond = - {\nabla_{\vect{y}} \vect{\xi}_\vect{\theta}\left (\vect{y}^\diamond\right )}^{-1} \nabla_{\vect{\theta} } \vect{\xi}_\vect{\theta}\left (\vect{y}^\diamond\right ) ,
\end{align}
where $\vect{\xi}_\vect{\theta}(\vect{y})$ gathers the primal-dual KKT conditions underlying the NMPC scheme \eqref{eq:param_nmpc}. We remind here that the component of $\vect{\xi}_\vect{\theta}(\vect{y})$ coincide with the gradient of the Lagrangian $\nabla_\vect{y} \mathcal{L}_\vect{\theta}(\vect{y})$ with the components corresponding to the inactive constraints removed. 
For a complete discussion on parametric sensitivity analysis of NLPs we refer to~\citep{Buskens2001} and references therein.

\subsection{RTI Sensitivities}

The sensitivity equations provided in the previous subsection are only valid if the NMPC problem is solved to full convergence. In case RTI or another approximate algorithm is used instead, the sensitivities must be computed differently. 
To that end, rather than considering RTI as a scheme approximately solving Problem~\eqref{eq:param_nmpc} or~\eqref{eq:param_nmpc_s}, it is best to view it as a scheme solving the QP Problem~\eqref{eq:param_nlp_qp} to compute $Q_\vect{\theta}$, $V_\vect{\theta}$, and $\vect{\pi}_\vect{\theta}$. Note that these function approximations depend on the initial guess used in the computation of the QP data. However, since RTI always uses a good initial guess which is close to the optimal solution, the impact of variations in the initial guess on the function approximations is small.

Analogously to the fully converged case, we denote the solution to~\eqref{eq:param_nlp_qp} as $\vect{y}^\star_\mathrm{QP}$, $\vect{y}^\diamond_\mathrm{QP}$ to respectively refer to the cases in which $\vect{u}_0=\vect{a}$ is enforced or not.
Then, based on the results provided above, the sensitivities are given by
%
%
%
%
\begin{align*}
\nabla_\vect{\theta} Q_\vect{\theta}(\vect{s},\vect{a}) &= \nabla_\vect{\theta} \mathcal{L}^\mathrm{QP}_\vect{\theta}(\vect{y}^\star_\mathrm{QP}), \\
\nabla_\vect{\theta} V_\vect{\theta}(\vect{s}) &= \nabla_\vect{\theta} \mathcal{L}^\mathrm{QP}_\vect{\theta}(\vect{y}^\diamond_\mathrm{QP}), \\
\nabla_{\vect{\theta}} \vect{y}^\diamond_\mathrm{QP} &= - {\nabla_{\vect{y}} \vect{\xi}^\mathrm{QP}_\vect{\theta}\left (\vect{y}_\mathrm{QP}^\diamond\right )}^{-1} \nabla_{\vect{\theta} } \vect{\xi}^\mathrm{QP}_\vect{\theta}\left (\vect{y}_\mathrm{QP}^\diamond\right ).
\end{align*}

We stress for completeness that, for $\vect{y}=\vect{\bar y}+\Delta\vect{y}$ and $L^{(i)}=\nabla^2_{\vect{z}\vect{z}} \mathcal{L}_\vect{\theta}\left (\vect{y}^{(i)}\right )$, the QP Lagrangian satisfies
\begin{align*}
	\mathcal{L}_\vect{\theta}^\mathrm{QP}(\vect{y}) &= \mathcal{L}_\vect{\theta}(\vect{\bar y}) +  \nabla_\vect{y} \mathcal{L}_\vect{\theta}(\vect{\bar y})^\top \Delta\vect{y} + \Delta\vect{y}^\top \nabla^2_\vect{yy} \mathcal{L}_\vect{\theta}(\vect{\bar y}) \Delta\vect{y} \\
	&= 	\mathcal{L}_\vect{\theta}(\vect{y}) + \mathcal{O}\left (\|\Delta \vect{y}\|^3\right ),
\end{align*}
such that
\begin{align*}
	\vect{\xi}_\vect{\theta}^\mathrm{QP}(\vect{y}) = \vect{\xi}_\vect{\theta}(\vect{\bar y}) + \nabla_\vect{y} \vect{\xi}_\vect{\theta}(\vect{\bar y})^\top \Delta\vect{y} +  \mathcal{O}\left (\|\Delta \vect{y}\|^2\right ).
\end{align*}


\section{RL based on RTI NMPC}
\label{sec:rl_mpc}

In this section, we revise the standard RL algorithms presented in Section~\ref{sec:background} and introduce some adaptations in order to account for the peculiarities of using NMPC as a function approximator. 

The update~\eqref{eq:Qlearning:Update} can also be written as the solution to a fitting problem. To that end, we introduce the linearization
\begin{align*}
	Q^\mathrm{lin}_\vect{\bar \theta}(\vect{s}_k,\vect{a}_k) = Q_\vect{\theta}(\vect{s}_k,\vect{a}_k) - \nabla_\vect{\theta}Q_\vect{ \theta}(\vect{s}_k,\vect{a}_k) (\vect{\bar \theta} - \vect{\theta}).
\end{align*}
Then,~\eqref{eq:Qlearning:Update} is equivalent to 
$\vect{\theta} \leftarrow \vect{\bar \theta}^*$, with $\vect{\bar \theta}^*$ computed as the optimal solution of the fitting problem
\begin{align}
	\label{eq:q_learning_fitting}
	 \min_{\vect{\bar \theta}} & \left (\ell(\vect{s}_k,\vect{a}_k) + \gamma V_\vect{\theta}(\vect{s}_{k+1}) - Q^\mathrm{lin}_\vect{\bar \theta}(\vect{s}_k,\vect{a}_k) \right )^2 + \frac{1}{\alpha}\left \|\vect{\bar \theta}-\vect{\theta}\right \|_2^2.
\end{align}

In order to guarantee that the NMPC cost is positive-definite, we propose to compute $\vect{\theta}^*$ as the solution of a slightly modified version of~\eqref{eq:q_learning_fitting}, i.e., 
\begin{subequations}
	\label{eq:q_solution}
	\begin{align}
	\hspace{-0.22em}\min_{\vect{\bar \theta}}  & 
	\left (\ell(\vect{s}_k,\vect{a}_k) + \gamma V_\vect{\theta}(\vect{s}_{k+1}) - Q^\mathrm{lin}_\vect{\bar \theta}(\vect{s}_k,\vect{a}_k) \right )^2 + \frac{1}{\alpha}\left \|\vect{\bar \theta}-\vect{\theta}\right \|_2^2
	\label{eq:q_solution_cost}\\
	\hspace{-0.22em}\mathrm{s.t.}  & \ \nabla^2 \ell_{\vect{\theta}} \succ 0, \quad \nabla^2 V^\mathrm{f}_{\vect{\theta}} \succ 0, \quad \forall \, \vect{s},\vect{a} \in \mathrm{dom}\{\eqref{eq:param_nmpc}\}. \label{eq:q_pd_constr}
	\end{align}
\end{subequations}
Constraint~\eqref{eq:q_pd_constr} imposes that the Hessian of the stage and terminal cost is positive-definite everywhere. We left this formulation intentionally implicit, since it can be enforced in several ways. For quadratic functions it can, e.g.,  be formulated as an LMI.
Alternatively, one can rely on parametrizations which deliver a positive-definite function by construction. Note that the positive-definiteness requirement is only necessary for those state-action pairs for which all NMPC problems have a solution. In principle, this set could be further restricted to the state-action pairs which will be visited when operating the system.

Additionally to the enforcement of positive-definiteness, as stressed in~\citep{Zanon2019}, the other main advantage of formulating $Q$-learning as a fitting problem is the possiblity to introduce globalization strategies such as, e.g., line search. This feature provides at least the guarantee that the proposed update $\Delta \vect{\theta}$ reduces the TD-error for the current sample, which is not the case with~\eqref{eq:Qlearning:Update}.

Similarly to $Q$-learning, for actor-critic methods, the parameter update~\eqref{eq:pg_update} can be replaced by $\vect{\theta} \leftarrow \vect{\theta}^*$ with $\vect{\theta}^*$ solution of
\begin{subequations}
	\label{eq:pg_solution}
	\begin{align}
		\min_{\vect{\theta}}  \ & \nabla_\vect{\theta} J(\vect{\pi}_\vect{\theta}(\vect{s}_k))^\top \hspace{-3pt} \left (\vect{\bar \theta}-\vect{\theta}\right ) + \frac{1}{\alpha}\left \|\vect{\bar \theta}-\vect{\theta}\right \|_2^2
		\label{eq:pg_solution_cost}\\
		\mathrm{s.t.} \ & \nabla^2 \ell_{\vect{\theta}} \succ 0, \quad \nabla^2 V^\mathrm{f}_{\vect{\theta}} \succ 0, \quad \forall \, \vect{s},\vect{a} \in \mathrm{dom}\{\eqref{eq:param_nmpc}\}. \label{eq:pg_pd_constr}
	\end{align}
\end{subequations}
with $\Delta \vect{\theta}^*=\vect{\bar \theta}^* - \vect{\theta}$. The considerations made for Problem~\eqref{eq:q_solution} also apply to Problem~\eqref{eq:pg_solution}.

\begin{Remark}
	Both in Problem~\eqref{eq:q_solution} and~\eqref{eq:pg_solution} one could use the nonlinear model of $Q$ and $J$ rather than the linear one. The thorough investigation of the effects of this choice are the subject of ongoing research.
\end{Remark}

\section{Simulations}
\label{sec:simulations}

In this section we demonstrate the effectiveness of the proposed combination of RL and RTI-based NMPC with an example from the process industry, i.e., the evaporation process modelled in~\citep{Wang1994,Sonntag2006} and used in~\citep{Amrit2013a,Zanon2016b} to demonstrate the potential of economic MPC in the nominal case. 
The model equations are given by 
\begin{align}
M \dot X_2 &= F_1 X_1 - F_2 X_2, &
C \dot P_2 &= F_4-F_5, 
\end{align}
where 
\begin{align}
T_2 &= aP_2 + bX_2 + c, & T_3 &= dP_2 + e, \nonumber \\
\lambda F_4 &= Q_{100} - F_1C_\mathrm{p}(T_2-T_1), & T_{100} &= fP_{100} + g, \nonumber
\end{align}
\begin{align}
 Q_{100} &= UA_1(T_{100}-T_2), & UA_1 &= h(F_1+F_3), \nonumber\\
 Q_{200} &= \frac{UA_2(T_3-T_{200})}{1+UA_2/(2C_\mathrm{p}F_{200})}, & F_{100} &=\frac{Q_{100}}{\lambda_\mathrm{s}} , \nonumber\\
\lambda F_5 &= Q_{200}, & F_2 &= F_1-F_4,\nonumber
\end{align}
with states $\vect{x} = (X_2, \, P_2)$ (concentration and pressure) and controls $\vect{u}=(P_{100},\,F_{200})$ (pressure and flow). The model parameters are given in~\citep{Amrit2013a}. 
The model further depends on concentration $X_1$, flow $F_2$, and temperatures $T_1,T_{200}$, which are assumed to be constant in the control model. In reality, these quantities are stochastic. In this example, we assume a uniform distribution around the nominal value with interval $\Delta_{X_1}=\pm1$, $\Delta_{F_1}=\pm2$, $\Delta_{T_1}=\pm8$, $\Delta_{T_{200}}=\pm5$. Additionally, the controller must satisfy bounds $(25,40)\leq (X_2,P_2) \leq (100,80)$ on the states and $100 \leq (P_{100},F_{200}) \leq 400$ on the controls. In particular, the bound $X_2\geq 25$ is introduced in order to ensure sufficient quality of the product.
The state bounds are relaxed as $(\vect{x}-\vect{x}_\mathrm{l}-\vect{\sigma},\vect{x}_\mathrm{u}-\vect{x}-\vect{\sigma})$, where $\vect{\sigma}$ is a slack variable introduced as a ficticious control and penalized in the cost in order to introduce an exact constraint relaxation~\citep{Scokaert1999a}. 
The stage cost is then given by 
\begin{align*}
\ell(\vect{x},\vect{u}) =\ &10.09(F_2+F_3) + 600 F_{100} + 0.6 F_{200} \\
&\hspace{12em} + \vect{\sigma}^\top B_\vect{\sigma}\vect{\sigma} + \vect{b}_\vect{\sigma}^\top \vect{\sigma}.
\end{align*}
For the given stage cost, the nominal model is optimally operated at the steady state $\vect{x}_\mathrm{s}=(25,49.74)$, $\vect{u}_\mathrm{s}=(191.71,215.89)$, with stage cost $\ell(\vect{x}_\mathrm{s},\vect{u}_\mathrm{s}) =\ell_\mathrm{s}$.

We parametrize an NMPC controller as in~\eqref{eq:param_nmpc}, i.e., a nonlinear MPC formulation, 
with $N=10$. Functions $\lambda_\vect{\theta}, V^\mathrm{f}_\vect{\theta}, \ell_\vect{\theta}$ are quadratic of the form
\begin{align*}
	\dagger=\matr{c}{\vect{x}-\vect{x}_\mathrm{s} \\ \vect{u}-\vect{u}_\mathrm{s}}^\top B_\dagger \matr{c}{\vect{x}-\vect{x}_\mathrm{s} \\ \vect{u}-\vect{u}_\mathrm{s}} + \vect{b}_\dagger^\top \matr{c}{\vect{x}-\vect{x}_\mathrm{s} \\ \vect{u}-\vect{u}_\mathrm{s}} + c_\dagger,
\end{align*}
 defined by Hessian $B_\dagger$, gradient $\vect{b}_\dagger$, and constant $c_\dagger$, with a minimum in $x_\mathrm{s}$, $u_\mathrm{s}$, and $\dagger=\{\lambda_\vect{\theta}, V^\mathrm{f}_\vect{\theta}, \ell_\vect{\theta}\}$. 
 The model is parametrized as the nominal model with the addition of a constant, i.e., $\vect{f}_\vect{\theta}(\vect{x},\vect{u}) = \vect{f}(\vect{x},\vect{u}) + \vect{c}_\vect{f}$. The control constraints are fixed and the state constraints are parametrized as simple bounds, i.e., $\vect{h}_\vect{\theta}(\vect{x},\vect{u})=(\vect{x}-\vect{x}_\mathrm{l}-\vect{\sigma},\vect{x}_\mathrm{u}-\vect{x}-\vect{\sigma})$.
The parameter vector reads as:
\begin{align*}
\vect{\theta} = ( B_\lambda, \vect{b}_\lambda, c_\lambda, B_{V^\mathrm{f}}, \vect{b}_{V^\mathrm{f}}, B_l, \vect{b}_l, \vect{c}_\vect{f}, \vect{x}_\mathrm{l},\vect{x}_\mathrm{u}).
\end{align*}
Constants $\vect{b}_\vect{\sigma}=10^5\vect{1}$, $B_\vect{\sigma}=I$ are fixed and assumed to reflect the known cost of violating the state constraints.

We apply $Q$-learning and use $\alpha=10^{-3}$. In order to induce enough exploration, we use an $\epsilon$-greedy policy which is greedy $90\, \%$ of the samples, while in the remaining $10 \, \%$ we apply the action
\begin{align*}
	\vect{a} = \mathrm{sat}(\vect{e},\vect{u}_\mathrm{l},\vect{u}_\mathrm{u}), && \vect{e} \sim \mathcal{N}(0,\sqrt{10}I),
\end{align*}
where $\mathrm{sat}(\cdot,\vect{u}_\mathrm{l},\vect{u}_\mathrm{u})$ saturates the input between its lower and upper bounds $\vect{u}_\mathrm{l},\vect{u}_\mathrm{u}$, respectively.

\begin{figure}
	\begin{center}
		\includegraphics[width=0.96\linewidth,clip,trim=0 145 0 100]{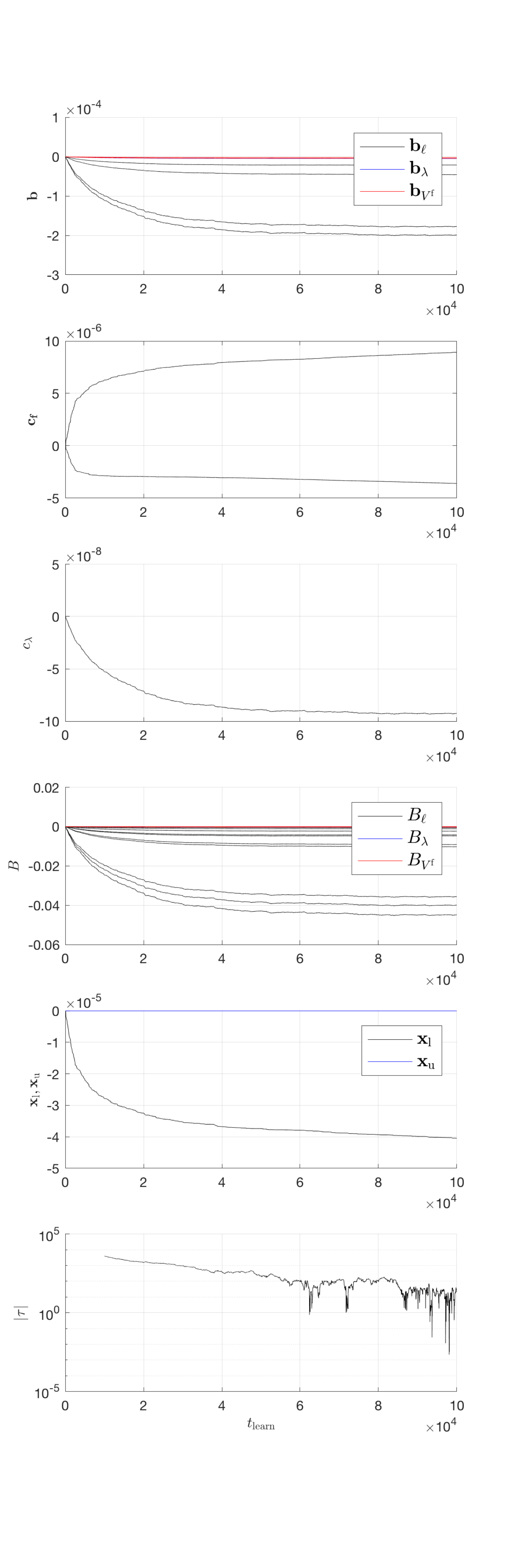}
	\end{center}
	\caption{Evolution of the parameters (increment w.r.t. the initial guess value) and of the TD error (averaged over the preceding $10000$ samples).}
	\label{fig:nmpc_learn}
\end{figure}

We initialize the ENMPC scheme by the 
naive initial guess $H_\ell=I$, $H_{V^\mathrm{f}}=I$, $c_\lambda=\ell_\mathrm{s}$, $\vect{x}_\mathrm{l}=(25,40)$, $\vect{x}_\mathrm{u}=(100,80)$, while all other parameters are $0$. 
As displayed in Figure~\ref{fig:nmpc_learn}, the algorithm converges to a constant parameter value while reducing the average TD-error. 

We performed a simulation to compare the RL-tuned RTI NMPC scheme to the same scheme using as parameter (a) the naive initial guess, and (b) the nominal model and the cost tuned using the economic-based approach proposed in~\citep{Zanon2016b}. The economic gain obtained by RL is approximately $21\,\%$ and $15\,\%$ respectively.

Note that when RL is deployed with fully converged NMPC as function approximator, the obtained optimal parameters are different from those obtained with RTI NMPC as function approximator.

\section{Conclusions and Outlook}
\label{sec:conclusions}

In this paper we have extended the framework of RL based on NMPC in order to make it possible to deploy computationally efficient algorithms for NMPC such as the RTI scheme and we have demonstrated the effectiveness of the approach with an example from the process industry.

Future research will aim at further developing the framework of MPC-based RL by providing a thorough stability analysis, as well further extending the analysis on rigorous safety guarantees started in~\citep{Zanon2019b,Gros2020d}.

\bibliography{my_bib}

\end{document}